\documentclass{hep99new}
\begin{document}

\title{QCD analysis of xF$_3$ at NNLO:\\
the theoretical uncertainties
}

\author{A.L. Kataev$^1$, G. Parente$^2$ and A.V. Sidorov$^3$}
%

\address{$^1$ Institute for Nuclear Research of the Academy of Sciences
of Russia,\\ 117312 Moscow, Russia\\
$^2$ Departamento de F\'\i sica de Part\'\i culas, Universidade de
Santiago de Compostela, Spain\\
15706 Santiago de Compostela\\
$^3$ Bogoliubov Laboratory of Theoretical Physics, Joint
Institute for Nuclear Research,\\ 141980 Dubna, Russia\\[3pt]
E-mails: {\tt gonzalo@fpaxp1.usc.es} }

\abstract{
The next-to-next-to-leading order (NNLO) QCD
analysis of the experimental $xF_3$ structure function from CCFR data
is performed. The theoretical uncertainties of the analysis are discussed.
}

\maketitle



The study of DIS structure functions has been
a fruitful source of information on the structure of the proton
and for testing perturbative QCD.
Between the most outstanding results of the program of
analyzing structure functions in QCD was the early explanation of the
scaling violation phenomena. In the last two decades the 
second order approximation (NLO)
has been extensively compared with
$F_2$ and $xF_3$ data. 
More recently the third order terms (NNLO) have been fully calculated
for the case of the coefficient functions \cite{WZ} but only
partially for the anomalous dimension part
($n = 2,4,6,8$ for singlet and non-singlet operators,
and $n=10$, only in the non-singlet case) \cite{LRV1}.
This has allowed the analysis of $xF_3$
\cite{KKPS1,KPS2} and $F_2$ (non singlet \cite{PKK} and singlet \cite{SY})
at NNLO.

In this note we review
the theoretical uncertainties involved in the
analysis of the structure function $xF_3$ at NNLO.
For that task, we firstly remind the method of
calculation and the most relevant results of the fits.


The QCD evolution of the moments is given by:
\begin{equation}
\frac{M_n(Q^2)}{M_n(Q_0^2)}=
\bigg(\frac{\alpha_s(Q^2)}{\alpha_s(Q_0^2)}\bigg)
^{\frac{\gamma_{NS}^{(0)}}{2\beta_0}}
\frac{AD(n,Q^2)C^{(n)}(Q^2)}{AD(n,Q_0^2)C^{(n)}(Q_0^2)}
\end{equation}
where $AD$ and $C$ come respectively from the
anomalous dimensions and coefficient function terms in the
renormalization-group equation (see the explicit forms in \cite{KPS2}).
The running coupling constant $\alpha_s(Q^2)$ is obtained from
the expression in terms of inverse powers of
$ln(Q^2/\Lambda_{\overline{MS}}^2)$.
Target mass corrections are also added in the calculation of
the moments (see \cite{KPS2}).

The moments in Eq. (1) at the initial scale are 
$M_n(Q_0^2)=\int_0^1 dx x^{n-2} A x^{b} (1-x)^{c} (1+\gamma x)$.

The structure function is reconstructed from its moments by
using the expansion in terms of orthogonal Jacobi polynomials:
\begin{equation}
xF_3
=  x^{\alpha}(1-x)^{\beta}\!\sum_{n=0}^{N_{max}}
\!\Theta_n^{\alpha,\beta}(x) 
\! \sum_{j=0}^{n}c_j^{(n)}(\alpha,\beta)
M_{j+2}(Q^2) 
\end{equation}
where  $c_j^{(n)}(\alpha,\beta)$ are combinatorial coefficients given in terms
of Euler $\Gamma$-functions of the $\alpha$ and $\beta$ weight parameters
which have been fixed to $0.7$ and $3$ respectively by the
reasons given in \cite{KPS2,ChR1}.

Power corrections are included in the
analysis using two different approaches. Firstly, in
the form given by the Infrared
Renormalon Model (IRR) \cite{DW} adding in Eq. (1) the contribution
$M_n^{IRR}=\tilde{C}(n)M_n(Q^2)A_2^{'}/Q^2$,
with $A_2^{'}$ a free scale parameter.
Secondly, adding in the r.h.s. of Eq. (2) the term
$h(x)/Q^2$, with $h(x)$ a free parameter for each $x$-bin of the data set.


Table 1 summarizes the results of the fits to xF$_3$ CCFR data \cite{CCFR}.
For comparison we have applied the same kinematic cuts as
in Ref. \cite{CCFR}, i.e. Q$^2$ $>$ 5 GeV$^2$, $x$
$<$ 0.7 and W$^2$ $>$ 10 GeV$^2$. At NLO the value of
$\Lambda_{\overline{MS}}^{(4)}$ from our fits
is in good agreement with that found in
Ref. \cite{CCFR} (337$\pm$28 MeV) where
both, F$_2$ and xF$_3$ structure functions, have been fitted.
There is a clear correlation between the effects of the NLO and NNLO
approximations and power corrections
(see table 1). At NNLO the fits performed with and without power corrections
(in the IRR model) are almost equal.
The significant decrease of the magnitude of power corrections
in the NNLO fit with IRR model ($A_2^{'}$ vanished within statistical
errors) is also found with the model $h(x)/Q^2$ (see Fig. 1).

Using the value of $\Lambda$ from the NNLO fit and the running
of the coupling up to $M_Z^2$, we obtain \cite{KPS2}
$\alpha_s(M_Z) = 0.118 \pm 0.002 (stat)\pm 0.005 (syst) \pm 0.003(theory)$.
The theoretical error takes into account
the dependence on the initial $Q_0^2$, the influence of
the missing higher order terms estimated by Pade approximants and
the crossing of the $m_b$ threshold in the calculation of $\alpha_s(M_Z)$.

However, in the analysis there are also involved
various approximations and shortcuts
which could increase this uncertainty.
The calculation of $xF_3$ with even-n $F_2$ anomalous dimensions
\cite{KKPS1}, the interpolation to odd values of $n$ \cite{PKK}
and the effect of the reconstruction method through the parameters
$\alpha$, $\beta$ and the number of polinomials $N_{max}$ (see Eq. (2))
\cite{KPS2,ChR1}, are not expected to affect the accuracy of the analysis.

In addition, we have also studied the effect of using in Eq. (1)
the original exponenciated formula for the anomalous dimension part
(see Eq. (4) in Ref. \cite{KPS2}). We found a change
in $\Lambda$ of 2 MeV at NLO and much smaller at NNLO.
The effect of nuclear corrections has also
been addressed by us \cite{KPS2}
although it still deserves a more detailed study. The dependence with
the number of active flavors (we work with $n_f$ = 4) should also be
carefully studied (see Ref. \cite{SY}).

Finally, the renormalization and factorization scale
dependence (we have fixed both equal to $Q^2$) should also be estimated
if one wants to make a meaningful precision test of perturbative QCD.
We plan to present this work elsewhere \cite{InProgress}.

\begin{table}
\begin{center}
\caption{Fits to $Q^2>5~GeV^2$ $xF_3$ CCFR'97 data.
The starting QCD evolution point is $Q_0^2=20~GeV^2$ }
\begin{tabular}{llll} 
\br
                    $  $                & 
 $\Lambda_{\overline{MS}}^{(4)}$ (MeV)  &
 $ A_2^\prime$(GeV$^2$)                 &
 $ \chi^2$/points                       \\
\mr
LO & 264$\pm$37      & --              & 113.1/86    \\
   & 433$\pm$53      & -0.33$\pm$0.06  & 83.1/86     \\
   & 331$\pm$162     & h(x) in Fig.1   & 66.3/86     \\ 
\mr
NLO & 339$\pm$42     & --             & 87.6/86      \\
    & 369$\pm$39     & -0.12$\pm$0.06 & 82.3/86      \\
    & 440$\pm$183    & h(x) in Fig.1  & 65.7/86      \\ 
\mr
NNLO & 326$\pm$35    & --             & 77.0/86      \\
     & 327$\pm$35    & -0.01$\pm$0.05 & 76.9/86      \\
     & 372$\pm$133   & h(x) in Fig.1  & 65.0/86      \\ 
\br
\end{tabular}
\end{center}
\end{table}

\input epsf
\begin{figure}[htp]
\centering

\vspace*{-5mm}
\hspace*{-5.5mm}
\leavevmode\epsfysize=8.5cm\epsfbox{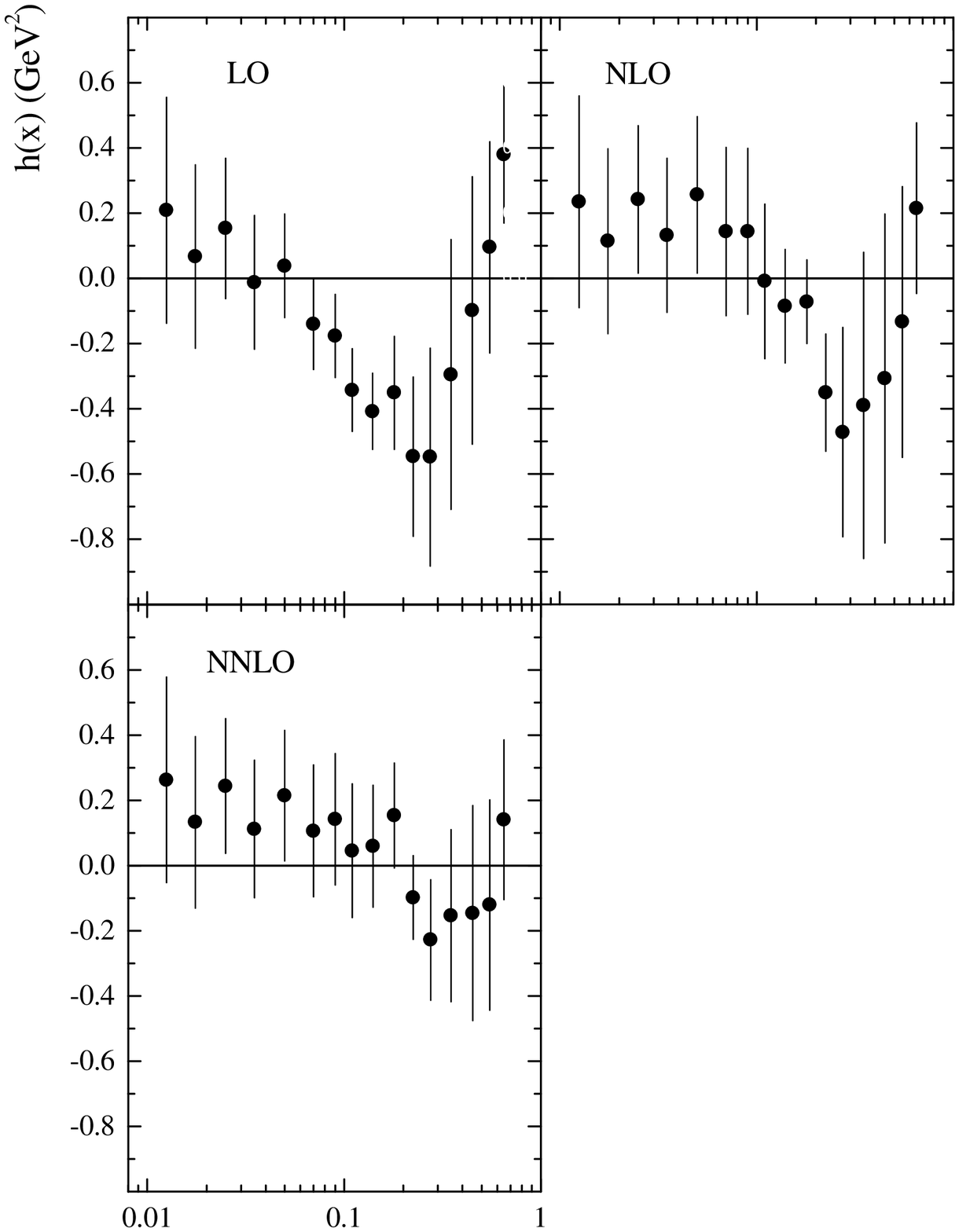}
\vspace*{-7mm}
\begin{center}
{\bf Figure 1.} $h(x)$ extracted from 
$xF_3$ CCFR data.
\end{center}
\end{figure}

\vskip 0.2cm

\noindent
{\bf Acknowledgments}
\vskip 0.2cm

G.P. is grateful to J. Ch\'yla for useful comments and
interest on this work.
A.L.K. and A.V.S. are supported by RFBR Grant N 99-01-00091.
The work of G.P. is supported by CICYT Grant N AEN96-1773
and Xunta de Galicia Grant N XUGA-20602B98.

\end{document}